\documentclass{aa}
\usepackage{graphicx}
\usepackage{txfonts}

\begin{document}
   
   \title{Dissipation of jet bulk kinetic energy in powerful blazars}
   
   %\subtitle{version 0.7}
   
   \author{Katarzy\'nski K. \inst{1,2} \& 
           Ghisellini G.    \inst{1}
	  }

   \offprints{Krzysztof Katarzy\'nski \\kat@astro.uni.torun.pl}

   \institute{Osservatorio Astronomico di Brera, via Bianchi 46, Merate and via Brera 28, 
              Milano, Italy  \and 
              Toru\'n Centre for Astronomy, Nicolaus Copernicus University, 
              ul. Gagarina 11, PL-87100 Toru\'n, Poland
             }

   \date{Received 26 September 2006 / Accepted 20 October 2006}

\abstract{}
{
We investigate the dissipation of the bulk kinetic energy of a relativistic jet
at different distances from the central power--house and analyse in detail
how the dissipated energy is radiated away.
}
{
We assume that the location of the dissipation region is a function of 
the bulk Lorentz factor $\Gamma$ of the jet, being closer to the center
for smaller $\Gamma$.
This assumption is naturally fulfilled in the internal shock scenario. 
The dissipated energy is partially used to accelerate electrons and to 
amplify the magnetic field. This process creates a source inside the
jet (blob). Such blobs may efficiently produce synchrotron and 
inverse Compton emission.}
{
We find that even if the blobs or shells responsible for the blazar
activity carry the same energy (in bulk kinetic form), the fact that
they move at different $\Gamma$ can produce dramatic variations
in different bands, even if the bolometric luminosity 
is instead very similar. 
This is due to the relative importance of the synchrotron,
self--Compton and external Compton radiation processes,
which greatly changes by changing $\Gamma$ and the compactness
of the source, even if the total radiated energy is constant.
We then find that the jet can produce most of its radiative output 
at small distances from the putative black--hole and its accretion 
disk, if this implies a low level of emitted MeV--GeV flux.
Our findings, which we apply for illustrative purposes to the blazar
3C 454.3, will be easily testable by the coming $\gamma$--ray satellite,
such as AGILE and GLAST.
}
{}

\titlerunning{Dissipation of bulk kinetic energy in powerful jets}
\authorrunning{Katarzy\'nski \& Ghisellini}   
\maketitle

\section{Introduction}

Our knowledge of the spectral energy distribution (SED) of blazars,
at MeV--GeV energies, has been built upon the results
of the EGRET instrument onboard the {\it Compton Gamma Ray Observatory},
which shows that blazars as a class are powerful $\gamma$--ray emitters,
and that the entire SED is characterized by two broad peaks,
with the high energy one becoming more dominant for more
powerful blazars (Fossati et al. \cite{Fossati98}).
This allows to define a blazar sequence, whose main
parameter is the bolometric observed luminosity,
dictating the overall spectral appearance of the SED
(Ghisellini et al. \cite{Ghisellini98}): in more powerful objects,
in which the radiative cooling is stronger, the emitting electrons have 
relatively small typical energies, explaining why in these sources the
peak frequencies of the two peaks are smaller than in low power blazars.
Since in these sources the stronger cooling is mainly due to the
presence of external radiation (e.g. the broad emission lines),
this scheme also explains why in powerful sources the high energy
peak dominates.
This blazar sequence has been obtained by averaging the data
of a relatively large number of blazars (more than one hundred),
and not all of them have been detected by EGRET.
It then accounts for the fact that a sizeable fraction
of blazars went undetected by EGRET.
On the other hand, this average does not entirely account for 
the fact that one specific source can spend most of the time
at MeV--GeV fluxes which are below the detection threshold.
Indeed, only a fraction (roughly 1/4) of bright radio blazars 
were detected by EGRET.
This can bias the representation of the average SED in the sense
that the high energy flux might better represents the 
high state of the source, rather than its time--average SED.

A related issue concerns the possibility of independent
variability of the two peaks.
Judging from the (few) examples we have, we know that 
variability at high energies (fluxes belonging to the high energy
peak) is accompanied with flux variability at smaller 
(i.e. UV, optical, IR, but not radio) frequencies 
(with the rare but interesting exceptions
of the so--called ``orphan flares", i.e. TeV flux variations 
not accompanied by simultaneous X--ray variations; 
Krawczynski et al. \cite{Krawczynski04}).

We then wonder if it is possible that a blazar flares in its
synchrotron part of the spectrum, but not in the MeV--GeV band.
As an illustrative example, consider the blazar 3C 454.3,  
detected by EGRET at the beginning of nineties 
(Hartman et al. \cite{Hartman99}), when the source was rather 
faint in the optical, and by {\it Beppo}SAX (in 2000, Tavecchio 
et al. \cite{Tavecchio02}). The {\it Beppo}SAX spectrum, up 
to $\sim 100$ keV, was relatively faint but very hard. 
Its extrapolation to the EGRET band was roughly consistent
with the flux detected by EGRET in a different period of time.
Recently, 3C 454.3 showed a huge flare in the optical (Fuhrmann 
et al. \cite{Fuhrmann06}, Pian et al. \cite{Pian06}) and in 
the X--ray bands, up to 100 keV, as detected by INTEGRAL and 
SWIFT (Pian et al. \cite{Pian06}; Giommi et al. \cite{Giommi06}). 
However, we have no information about the emission level in the 
MeV-GeV range during above mentioned activity.
The SED of 3C 454.3 was interpreted by Ghisellini et al. 
(\cite{Ghisellini98}) and Tavecchio et al. (\cite{Tavecchio02})
as due to a synchrotron and Inverse Compton model, produced by 
one--zone region of the relativistic jet, in analogy with other 
powerful blazars.
The high energy peak was the result of the inverse Compton scattering
of relativistic electrons off seed photons coming from the broad line
region (BLR), with the synchrotron self--Compton process playing 
a negligible role.
If we apply the very same model to the optical/X--ray flaring state
observed recently, then we are led to conclude that also the MeV--GeV
emission should show a flare as well, of similar amplitude to what 
seen in the optical
\footnote{If the observed flux variation is due to an increase of the total
number of electrons, then we expect, in this model, that variations
of the synchrotron and of the external Compton components are linear.}.
We are then led to conclude that the jet has indeed changed its total
power, by a large factor. Unfortunately, this scenario cannot be 
tested immediately due to lack of the MeV-GeV observations.
There are, however, more ``economic" solutions, in which the total
jet power changes much less, or it is even constant.
One solution was proposed in Pian et al. (\cite{Pian06}): in this paper it was
proposed that the emission site, responsible for the flare observed
in 2005, was {\it outside} the BLR, where the number of external photons
was negligible.
In this case the external Compton component becomes negligible,
and the power is mainly emitted by the synchrotron and self--Compton
process.
With roughly the same jet power, we can account both for the ``high EGRET"
state (with a large dominance of the high energy peak over the synchrotron
one) and the ``high synchrotron" state, where the power in the synchrotron
and synchrotron--self Compton components are roughly equal, and equal to the
previously observed ``high EGRET" state.

In this paper we explore yet another possibility, in which the emission
site is {\it within} the BLR, but the importance of the external Compton
component is much reduced because of a reduced bulk Lorentz factor $\Gamma$.
In fact, the importance of this component depends on the radiation energy
density $U^\prime_{\rm ext}$ of the seed external photons as seen in 
the comoving frame, and $U^\prime_{\rm ext}\propto \Gamma^2$.
It is then possible, in principle, to reduce the external Compton component
and at the same time to enhance both the synchrotron and the self--Compton ones.
To be more quantitative, we have also to specify not only the $\Gamma$--factor,
but also the size of the region, the magnetic field value, the density
of the particles and their energy spectrum.
There is a specific scenario where all these quantities are all
dependent in a specific way to the value of $\Gamma$, and this is 
the internal shock model.
We will discuss it in detail in the next section, but the basic idea
is that in this scenario the central engine is working intermittently,
producing shells of matter moving at different speeds (and hence different
Lorentz factors).
At the start they are separated by same distance $\Delta R$, and there
is always some probability that a later shell is faster then the previous
one. The fastest shell will then catch up the slower one, at a distance
of the order of $R_{\rm coll}\sim \Gamma^2 \Delta R$.
Therefore, if, on average, the two shells are produced with $\Gamma$
factors that are different, but relatively small, $R_{\rm coll}$
will be correspondingly small and the emission region will be more compact,
and the particle density and the magnetic field will be greater.
If, in addition, we also assume that the kinetic and magnetic energy
carried by the shells is always the same and that the
dissipated energy (in the collision) is also the same, we
have that all quantities depends on $\Gamma$ only.

The internal shock scenario then offers a nice possibility to
explore and quantify the consequences of the idea that a
typical relativistic jet, even if on average produces blobs
of equal energies and therefore works, on average, at a
constant power, yet it can produce spectacular flares in
different energy bands.
However, we would like to stress that the main idea explored 
in this paper is not to test the internal shock scenario,
but to investigate what happens if the jet, carrying on average
the same amount of energy, dissipates it at different distances
from the black hole, characterized by different $\Gamma$--factors
(smaller $\Gamma$ at shorter distances).
Our study also allows us to answer a more general question:
Is it mandatory that the jet always produces most of
the radiation we see at relatively large distances from 
the black hole?
If the jet dissipation always results in the production of 
a dominant MeV--GeV component the answer is yes.
As discussed in Ghisellini \& Madau (\cite{Ghisellini96}), if high energy
photons are produced too close to the accretion disk
and its X--ray corona, they would be absorbed by photon--photon
collisions, producing electron--positron pairs.
These pairs would Compton scatter the UV--optical radiation 
from the disk, reprocessing the energy originally in the MeV--GeV 
band into the X--ray band.
The observed ``valley" between the two broad peaks of the blazar SED
would then be filled by this reprocessed radiation.
But what happens if the jet, close to the disk {\it does not} produce
a too prominent MeV--GeV component?

\section{The model}

In our model we assume a relativistic jet that contains electrons, 
protons and tangled magnetic field. 
Different sub--structures of this 
jet (hereafter shells) may travel with different velocities. Therefore, 
some of the shells may collide and generate relativistic shock waves. 
The shock, through the first order Fermi acceleration, may increase 
significantly the energy of the electrons. 
In other words, a fraction of 
the bulk kinetic energy of the colliding shells is used
to accelerate electrons up to highly relativistic energies. 
The relativistic electrons escape from the shock front into the
downstream region of the shock, where they lose their energy 
through the synchrotron and the inverse-Compton processes. 
These are 
the main assumptions of the so--called internal shock scenario,
proposed for the first time by Rees (\cite{Rees78}) and applied 
successfully for gamma--ray bursts 
(see e.g. Meszaros 2006 for a recent review) 
and blazars (Sikora et al. \cite{Sikora94}, 
Begelman et al. \cite{Begelman94}; Ghisellini \cite{Ghisellini99}; 
Spada et al. \cite{Spada01}, Guetta et al. 2004). 

\subsection{Set up of the model}

Constructing our model we make two important assumptions.
The first concerns the efficiency of the central engine 
that forms and accelerates the shells. 
We simply assume that {\it the efficiency of the central engine 
is constant}. 
This assumption {\it is not} a key ingredient of our model,
but it allows to see if {\it even} if the jet is working
with the same efficiency, we can account for large variation
of the flux in different bands, as observed.
According to this first assumption, the bulk Lorentz factor 
$\Gamma=1/\sqrt{1-\beta^2}$ of the shell is 
inversely proportional to the total mass of the shell, so that the 
total energy given to the shell by the central engine is constant:
\begin{equation}
M \Gamma = {\rm const.},
\end{equation}
where $M \simeq N_p m_p$ ($N_p$ is total number of protons of 
mass $m_p$).

Consider now a collision of two shells ($a$ and $b$) of bulk 
kinetic energy $\Gamma_a M_a$ and $\Gamma_b M_b$ respectively. 
The collision must satisfy energy
\begin{equation}
\Gamma_a M_a + \Gamma_b M_b = \Gamma_s (M_a + M_b + E'/c^2),
\label{equ_energy_con}
\end{equation}
and momentum
\begin{equation}
\Gamma_a \beta_a M_a + \Gamma_b \beta_b M_b = \Gamma_s \beta_s 
(M_a + M_b + E'/c^2),
\label{equ_moment_con}
\end{equation}
conservation, where $\Gamma_s$, $\beta_s$ are 
describing the ``merged" shell (s) created from the colliding 
shells and $E'$ is the energy dissipated during the collision, 
measured in the comoving frame of the new shell. 
As mentioned above, this energy can be used to accelerate the 
electrons, and it can also be used for heating the
protons or to amplify the magnetic field intensity. 
Since it is the new shell which contains the relativistic
emitting particles, hereafter we will call it
simply ``source" or  ``blob". 
The  efficiency describing
how much of the kinetic energy of the two colliding shells is 
dissipated can be defined as
\begin{equation}
\eta = \frac{E}{\Gamma_a M_a + \Gamma_b M_b},
\end{equation}
where $E$ is the dissipated energy in the observer's frame.
Using the conservation laws (Eq. \ref{equ_energy_con} and Eq.
\ref{equ_moment_con}) we can derive
\begin{equation}
\eta = 1 - \frac{\Gamma_s (1+\alpha_M)}{\Gamma_a(1+\alpha_{\Gamma} 
\alpha_{M})},
\label{equ_effic}
\end{equation}
where $\alpha_M=M_b/M_a$ and $\alpha_{\Gamma}=\Gamma_b/\Gamma_a$.
Since our first assumption (shells of constant bulk kinetic energy)
implies  $\alpha_{\Gamma} = 1/\alpha_{M}$, the above formula 
simplifies into
\begin{equation}
\eta \simeq 1 - \frac{1+\alpha_M}{\sqrt{2+2\alpha^2_m}},
\end{equation}
(see e.g. Lazzati et al. \cite{Lazzati99}). 
This simple relation shows that $\eta$
varies between 0 and 22\% if the contrast between 
the Lorentz factors ($\alpha_{\Gamma}$) changes from 1 to 10.

Our second assumption concerns the amount of the energy 
transferred to the electrons during the collision. 
We simply assume that {\it the electrons receive always the 
same amount of energy} ($E_e={\rm const.})$.
Since our first assumption implies constant energy of the colliding
shells, the present assumption requires only that the contrast between the 
Lorentz factors of the colliding shells is always the same 
($\alpha_{\Gamma} = {\rm const.}$). 
We are aware that this may be a great simplification of what we can
expect in reality. 
However, we will see that this simple assumption is enough to 
provide a reasonable explanation for the some
properties of blazar variability. 

The total energy transferred to the electrons is given by
\begin{equation}
E_e = \Gamma E'_e = \Gamma N_e <\gamma> m_e c^2 = {\rm const.},
\label{equ_elec_energy}
\end{equation}
where $N_e$ is the total number of electrons and
\begin{equation}
<\gamma> = \frac{\int^{\gamma_{\rm max}}_{\gamma_{\rm min}} \gamma Q(\gamma) {\rm d} \gamma}
                {\int^{\gamma_{\rm max}}_{\gamma_{\rm min}} Q(\gamma) {\rm d} \gamma}
\end{equation}
is the average electron energy, where $\gamma=1/\sqrt{1-\beta_e^2}$
is the electron random Lorentz factor and $Q(\gamma)$ describes 
the injection of the particles in the energy range constrained 
by $\gamma_{\rm min}$ and $\gamma_{\rm max}$. 
In the model we describe the particle acceleration 
as a continuous injection of relativistic electrons. 
This injection is simulating the escape of the particles from the shock 
region into the downstream region of the shock that forms the source. 
First, we will discuss the very simple case of a monoenergetic 
injection, later we will discuss more complex, but more realistic, 
injections of power law energy distributions. 
The duration of the injection is assumed to be equal to 
the source light crossing time ($R/c$). 
During the process the electrons that are carrying the total energy 
described by Eq. \ref{equ_elec_energy} are injected into the source. 
The evolution of the particle energy distribution $N(\gamma, t)$ inside the source 
is calculated from the kinetic equation:
\begin{eqnarray}
\frac{\partial N(\gamma, t)}{ \partial t} 
& = & \frac{\partial }{\partial \gamma}
\left\{ C  \gamma^2N(\gamma, t)\right\}+ Q(\gamma),\\
C & = & \frac{4}{3} \frac{\sigma_{\rm T} c}{m_e c^2} 
(U'_B + U'_{\rm syn} + U'_{\rm ext}),
\end{eqnarray}
where $U'_B, U'_{\rm syn}, U'_{\rm ext}$ are the magnetic, 
synchrotron and external radiation field energy densities, respectively.
The solutions of this equation for the different types of  
injections, that we use in our modelling, were first
derived by Kardashev (\cite{Kardashev62}).

For the sake of simplicity, the source is approximated as a spherical 
homogeneous blob, of radius $R$ assumed to be proportional to the distance $D$
to the center of the active nuclei ($R =\psi D$, where $\psi$ 
is the jet aperture angle).
We also assume that the $D$ depends on the bulk Lorentz factor 
$\Gamma$ as $D \propto \Gamma^2$.
This assumption is supported by the internal shock model where
two shells ($\Gamma_b>\Gamma_a$) will collide when the second
source ($\Gamma_b$) will travel the distance
\begin{equation}
D = 2 \frac{\alpha^2_{\Gamma}}{\alpha^2_{\Gamma}-1} D_0 \Gamma^2_a,
\end{equation}
where $D_0$ is the initial separation of the sources 
(see e.g. Lazzati et al. \cite{Lazzati99}). 
In other words, relatively slow shells ($\Gamma \ll 10$) 
may collide closer to the center than faster shells 
($\Gamma \gtrsim 10$) that can collide only at larger distances.
Note that this formula is valid only for constantly moving shells
(no acceleration or deceleration). 
Moreover, this formula shows that our assumption ($D \propto\Gamma^2$)
becomes inaccurate for small $\Gamma$--contrasts ($\alpha_{\Gamma} < 2$). 
Finally, our assumption is supported by the fact that the
central engine is probably not able to accelerate almost 
instantaneously the jet components up to very high values of 
$\Gamma$. 
If the acceleration process requires a time comparable 
to the travel time of the central engine (e.g. 0.1 pc) then
we should expect fast shells farther from the center 
than slow shells.

To complete the general description of our model we have to define the
evolution of the magnetic field. 
By analogy to the definition of the energy transferred to the electrons 
(Eq. \ref{equ_elec_energy}), we assume that the total energy 
accumulated after the collision in the magnetic field, measured in 
the observer's frame, is constant
\begin{equation}
E_B = \Gamma E'_B = \Gamma U'_B V = \Gamma B'^2 V' / (8 \pi) = 
{\rm const.},
\end{equation}
where $B'$ is the magnetic field intensity and $V'$ is the source volume
as measured in the comoving frame. In other words, the fraction of the total 
dissipated energy used for the amplification of the magnetic field is constant.

Finally, some part of the dissipated energy can be transferred to
the protons. However, we neglect this process in our computations since 
protons are not assumed to radiate and since their energy 
changes slightly for different Lorentz factors.

\subsection{Comparison of different sources}

Our main aim is to investigate how the energy transferred to 
the electrons (that is constrained to be exactly the same for 
every source) is radiated by different sources. 
By ``different sources" we here mean different states of the same
jet.
Therefore, we simply compare the emission of two sources with 
different Lorentz factors ($\Gamma_2<\Gamma_1$). 
The difference in the Lorentz factors means that the sources 
were created by collisions of a pair of shells with different 
velocities (e.g. $\Gamma_b=20$ and 
$\Gamma_a=10 \to \Gamma_1 \simeq 13.3$ and $\Gamma_c=8$ and~~ 
$\Gamma_d=4 \to \Gamma_2 \simeq 5.3$). 
Note that according 
to our assumption the contrast between the Lorentz factors of 
the colliding shells ($\alpha_{\Gamma}=\Gamma_b/\Gamma_a=
\Gamma_c/\Gamma_d$) is constant. 
Moreover, we assume that the shells before the 
collision contain only ``cold'' particles that cannot emit. 
Therefore, it is not mandatory to describe the physical parameters 
of both sources:
it is sufficient to constrain the physical parameters of the first source 
using observations of the first state and derive the parameters of the 
second source (second state) choosing $\Gamma_2$ and using our 
assumptions
\begin{equation}
R'_2 = R'_1 \left(\frac{\Gamma_2}{\Gamma_1}\right)^2,
B'_2 = B'_1 \left(\frac{\Gamma_2}{\Gamma_1}\right)^{-3.5},
<\gamma>_2~ =~ \frac{N_{e,1} <\gamma>_1}{N_{e,2}}.
\end{equation}
Finally, to complete the description of the sources it is 
necessary to describe the injection function.

\subsection{Monoenergetic injection}

In the first approach we assume continuous monoenergetic 
injection of relativistic electrons
\begin{equation}
Q(\gamma) = Q_i \, {\rm Dirac} (\gamma - \gamma_i),
\end{equation}
where $Q_i$ is the number of injected particles per time unit and
$\gamma_i$ describes the energy of the injected electrons. 
This approach helps to simplify the description of the model 
(e.g. $<\gamma>\equiv \gamma_i$) and allows to derive simple formulas 
that describe the evolution of the observed emission. 
According to the solution of the kinetic equation, 
the $N(\gamma$) distribution is a power law 
\begin{equation}
N(\gamma) = \frac{Q_i}{C} \gamma^{-n}~~ 
{\rm for}~~ \gamma_{\rm min} \le \gamma \le \gamma_i; ~~
\gamma_{\rm min} = \frac{\gamma_i}{1+C \gamma_i T},
\label{equ_mono_inj_elec_spec}
\end{equation}
where $n=2$ in this particular case, $T'=R/c$ is the duration 
time of the injection assumed to be equal to the source light
crossing time. 
The total number of injected electrons is
\begin{equation}
N_e = V' T' \int Q(\gamma) {\rm d} \gamma,
\end{equation}
that gives $N_e = V' T' Q_i$ in the particular case of a 
monoenergetic injection. 
We assume similar number of electrons and protons inside the shells 
before the collision (e.g. $N_{e,a} \simeq N_{p,a} 
\neq N_{e,b} \simeq N_{p,b}$). 
However, only some fraction of the electrons may be accelerated. 
Moreover, we assume that the total number of accelerated electrons 
(total number of the injected electrons) is proportional to the total 
number of protons inside the source. 
According to our main assumptions 
($\Gamma_a M_a = \Gamma_b M_b = \Gamma_c M_c = 
\Gamma_d M_d = {\rm const.}$ and $\Gamma_1 E'_1 = 
\Gamma_2 E'_2 = {\rm const.}$) we obtain $\Gamma_1(M_a+M_b) 
= \Gamma_2 (M_c+M_d)$. 
If we neglect the contribution of the electrons to the total mass 
of the shells then the total number of protons and therefore also 
the total number of electrons in different sources scales as
\begin{equation}
N_{2} = N_{1} \frac{\Gamma_1}{\Gamma_2}.
\end{equation}
This gives
\begin{equation}
Q_{i, 2} = Q_{i,1} \frac{V'_1 T'_1 \Gamma_1}{V'_2 T'_2 \Gamma_2} = 
           Q_{i,1} \left( \frac{\Gamma_2}{\Gamma_1} \right)^{-9},
\end{equation}
where the injection time ($T'$) and the source volume ($V'$) 
scale as
\begin{equation}
T'_2 = T'_1 \left( \frac{\Gamma_2}{\Gamma_1} \right)^{2}, ~~~
V'_2 = V'_1 \left( \frac{\Gamma_2}{\Gamma_1} \right)^{6}.
\end{equation}
Moreover, since $\gamma_i \equiv <\gamma>$ the transformation
of the total number of the particles gives
$\gamma_{i, 2} = \gamma_{i, 1} \Gamma_2 / \Gamma_1$. The electron
density inside the source after the end of the monoenergetic
injection is defined by
\begin{equation}
K = \frac{Q_i}{C} = \, {3 m_e c^2 \over 4 \sigma_T c} \,
  {Q_i \over U'_B + U'_{\rm syn} + U'_{\rm ext} },
\label{equ_part_dens}
\end{equation}
where the magnetic field energy density ($U'_B$) determines the
synchrotron cooling during the injection, the synchrotron 
radiation field energy density ($U'_{\rm syn}$) determines the
cooling due to synchrotron self--Compton emission (this is
correct only in the Thomson limit) and the
radiation field energy density of the external medium, 
measured in the comoving frame of a source ($U'_{\rm ext}$) 
determines the cooling due to inverse--Compton scattering of 
external photons. 
We can relatively easily describe the change 
of $U'_B$ and $U'_{\rm ext}$ as a function of $\Gamma$.
However, the precise description $U'_{\rm syn}$ 
is quite complex and requires a numerical approach. 
Therefore, for deriving some analytical formulas describing the 
emission processes, we will focus on two simple and opposite
cases.

First, we assume the dominance of the EIC scattering 
($U'_{\rm ext} \gg U'_B$  and $U'_{\rm ext} \gg U'_{\rm syn}$). 
Moreover, $U'_{\rm ext}$ is assumed to be constant in the 
observer's frame with a good approximation if this external
radiation is produced by the Broad Line Region (BLR) 
(see e.g. Ghisellini \& Madau \cite{Ghisellini96}). 
Therefore, the external radiation field density in the 
comoving frame for sources with different Lorentz factors 
scales as
\begin{equation}
U'_{\rm ext, 2} = U'_{\rm ext, 1} \left( \frac{\Gamma_2}{\Gamma_1} \right)^2 
\end{equation}
This gives the following transformation of the particle density
\begin{equation}
K_{2} = K_{1} \left( \frac{\Gamma_2}{\Gamma_1} \right)^{-11},
\label{equ_part_dens_Uext}
\end{equation}

Knowing how all physical parameters of the source 
($R'$, $B'$, $K$, $\Gamma$, $U'_{\rm ext}$) scale with $\Gamma$, 
we can derive 
the approximated formulae describing the different types of 
emission, and how these depend on the bulk Lorentz factor.

First, consider the synchrotron emission where the emissivity at 
given frequency depends on the particle density and the magnetic 
field intensity as
\begin{equation}
j'_{\rm syn}(\nu') \propto K B'^{1+\alpha}\nu'^{-\alpha},
\end{equation}
where $\alpha=(n-1)/2$ is the energy spectral index.
In the case of mono--energetic injection, $\alpha=0.5$.
The observed thin synchrotron flux is given by
\begin{equation}
F_{\rm syn}(\nu) \propto 
\delta^{3+\alpha} R^2 I^\prime_{\rm syn}(\nu'),
\end{equation}
where $\delta = \left[ \Gamma (1-\beta \cos \theta) \right]^{-1}$ is
the source Doppler factor, $\theta$ is the viewing angle and
$ I^\prime_{\rm syn}(\nu^\prime) 
\propto R' j^\prime_{\rm syn}(\nu^\prime) 
\propto R' K B'^{1+\alpha}\nu '^{-\alpha}$ 
is the intensity of the thin synchrotron emission 
as measured in the comoving frame.
For simplicity we here neglect the self absorption regime
(we will mainly investigate the synchrotron emission in the optical 
and X--ray energy bands where self--absorption is negligible). 
Moreover, in 
order to simplify the description even more we assume $\Gamma=\delta$.
This is strictly true only if the viewing angle $\theta=1/\Gamma$ [rad]. 
This assumption enables possibility to describe the flux as a 
function of the Lorentz factor only. 
Strictly speaking this implies that when we compare the fluxes 
produced by sources with different Lorentz factors we compare 
these sources in different astrophysical objects with different 
viewing angles. 
However, this approximation does not change qualitatively our 
results in comparison to the case where we fix the viewing angle 
(this is mandatory when we study the same astrophysical object) 
and modify the value of the Doppler factor. 
Note that for $\theta=0$ the Doppler factor $\delta \sim 2 \Gamma$. 
We use this approach only for the tests presented
in this subsection for illustrating the relations between
different types of the source emission in a simpler way.
Using the above assumptions we can write 
$F_{\rm syn}(\nu) \propto \Gamma^{3.5} R^3 K B'^{1.5}\nu^{-0.5}$ 
that gives the following scaling
\begin{equation}
F_{\rm syn, 2}(\nu) = F_{\rm syn, 1} (\nu)
\left( \frac{\Gamma_2}{\Gamma_1} \right)^{-6.75}.
\end{equation}
This shows that the synchrotron specific flux increases
when $\Gamma$ decreases. 
This is because the increase of the magnetic field intensity and 
the particle density can easily compensate the decrease of source 
volume and the decreased Doppler boosting. 

In a similar way we can describe the evolution of the self--Compton
emission where the emissivity depends on the particle density and
the intensity of the radiation field 
$j'_{\rm SSC}(\nu) \propto K I'_{\rm syn}(\nu')$:
\begin{equation}
j'_{\rm SSC}(\nu') \propto K I'_{\rm syn}(\nu') 
\propto R' K^2 B'^{1.5}\nu'^{-0.5},
\end{equation}
The observed flux is
$ F_{\rm SSC}(\nu) \propto  \Gamma^{3.5} R^4 K^2 B'^{1.5}\nu^{-0.5}$,
which according to our previous assumptions scales as
\begin{equation}
F_{\rm SSC, 2}(\nu) = F_{\rm SSC, 1}(\nu)
 \left( \frac{\Gamma_2}{\Gamma_1} \right)^{-15.75},
\end{equation}
Note the stronger dependence on $\Gamma$ with respect to the
synchrotron flux.

Finally, the external inverse--Compton (EIC) flux is given by
$F_{\rm EIC}(\nu) = F_{\rm EIC,0} \nu^{-0.5} 
\propto \Gamma^{3.5} R^3 K I'_{\rm ext} \nu'^{-0.5}$. 
The (frequency integrated)
intensity of the external radiation field in the comoving
frame ($I_{\rm ext}$) is proportional to $\Gamma^{1.5}$. 
To understand this, assume a source with a constant size, 
and particle density, surrounded by a constant 
external radiation field. Change only $\Gamma$. 
The total flux ($F^{\rm tot}_{\rm EIC}$) 
produced by such a source is proportional to $\Gamma^2 \delta^4$, and,
according to our assumption ($\Gamma=\delta$),
$F^{\rm tot}_{\rm EIC} \propto \Gamma^6$. 
Assuming a power law spectrum of the emission we can write 
\begin{equation}
F^{\rm tot}_{\rm EIC} = 
\int^{\nu_{\rm max}}_{\nu_{\rm min}} F_{\rm EIC,0} \nu^{-0.5} d \nu 
\simeq F_{\rm EIC,0} \nu^{0.5}_{\rm max}.
\end{equation}
Since $\nu_{\rm max} \propto \delta\Gamma\sim \Gamma^2$, 
this relation gives $F_{\rm EIC,0} \propto \Gamma^5$. 
Moreover, for the source where $R$ and $K$ do not depend on
$\Gamma$ we have $F_{\rm EIC}(\nu) 
\propto \Gamma^{3.5} I'_{\rm ext}\nu'^{-0.5} \propto \Gamma^5$ 
that gives $I'_{\rm ext} (\nu') \propto \Gamma^{1.5}$.
Using this relation and the previous definitions we obtain
\begin{equation}
F_{\rm EIC, 2}(\nu) = F_{\rm EIC, 1}(\nu),
\end{equation}
as long as we compare the fluxes at a given frequency in the power law 
part of the spectrum. 
This means that our two sources with different Lorentz
factors have the same specific flux 
if the total energy transferred to the particles is the same 
for each source and the physical parameters of the sources are
scaling according to our prescriptions. 
This result is valid only for the sources where the 
emission is dominated by the external inverse--Compton scattering.

Consider now the opposite case where the cooling is dominated
by the synchrotron emission ($U'_B \gg U'_{\rm ext}$ and $U'_B \gg 
U'_{\rm syn}$). The magnetic field energy density transforms as
\begin{equation}
U'_{B, 2} = U'_{B, 1} \left( \frac{\Gamma_2}{\Gamma_1} \right)^{-7},
\end{equation}
that according to Eq. \ref{equ_part_dens} gives
\begin{equation}
K_{2} = K_{1} \left( \frac{\Gamma_2}{\Gamma_1} \right)^{-2}.
\end{equation}
This scaling differs significantly from the result obtained in the previous 
case (Eq. \ref{equ_part_dens_Uext}), where $U'_{\rm ext}$ was dominant. 
Therefore, all formulae describing the different types of the emission 
are now completely different. 
The synchrotron and SSC emission scale in the same way
\begin{eqnarray}
F_{\rm syn, 2}(\nu) &=& F_{\rm syn, 1}(\nu) 
\left( \frac{\Gamma_2}{\Gamma_1} \right)^{2.25}\nonumber \\
F_{\rm SSC, 2}(\nu) &=& F_{\rm SSC, 1}(\nu) 
\left( \frac{\Gamma_2}{\Gamma_1} \right)^{2.25}
\end{eqnarray}
this implies a decrease of the flux when $\Gamma$ is decreasing.
Also the EIC emission is decreasing with decreasing $\Gamma$
\begin{equation}
F_{\rm ext, 2}(\nu) = F_{\rm ext, 1}(\nu) 
\left( \frac{\Gamma_2}{\Gamma_1} \right)^{9}.
\end{equation}
Then, when the synchrotron cooling is dominant, all three radiative 
processes produce an observed flux which decreases with decreasing $\Gamma$.
Note that, in this context, a decrease of $\Gamma$ implies a comparison 
of two states of the jet, not a deceleration of a single source.

\begin{figure}[!t]
\resizebox{\hsize}{!}{\includegraphics{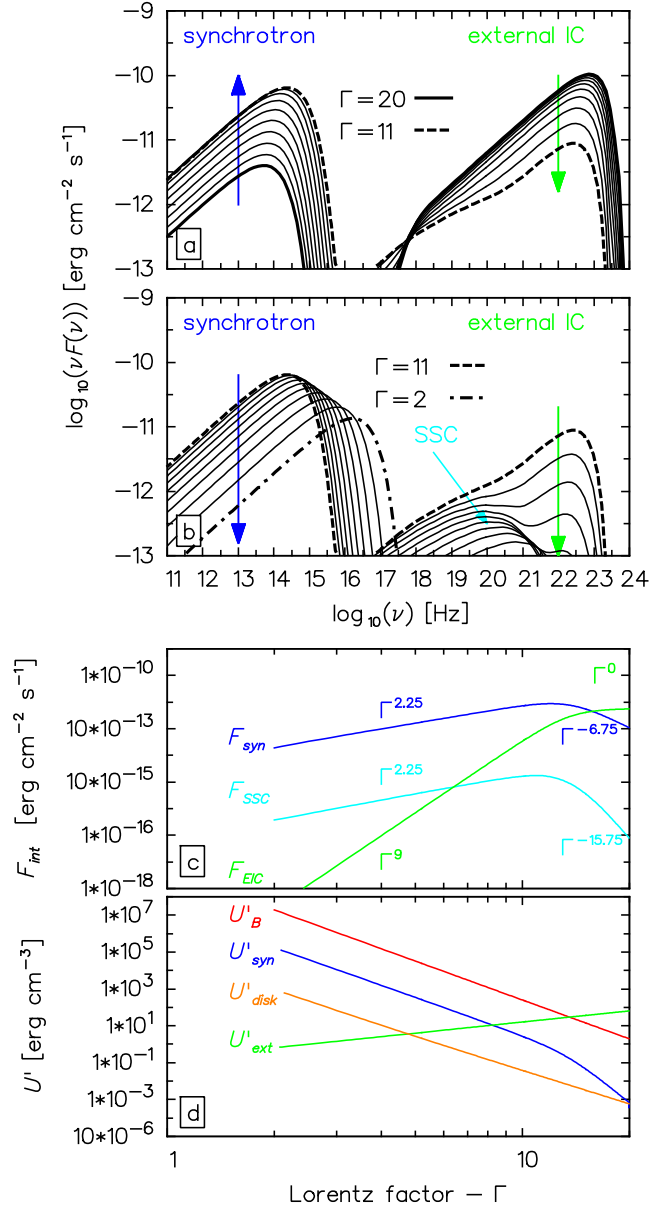}}
%\begin{figure}[!t]
%\includegraphics[height=19.5cm]{ps/art_mi1n_cp.ps} % fort the referee form
\caption{In the upper panels we compare total emission of sources 
         located at different distances from the center and
         travelling with different bulk Lorentz factors ($\Gamma=20 \to 11$ 
         panel a, and
         $\Gamma=11 \to 2$ panel b). The total
         energy transferred to the electrons is exactly the same for each 
         source. In this particular case the electrons are injected with a
         monoenergetic distribution. Panel c) shows, separately, the fluxes
         of the different emission mechanisms, integrated over narrow 
         frequency ranges: 
         synchrotron ($10^{11}$--$10^{12}$ Hz), SSC ($10^{14}$--$10^{15}$ 
         Hz) and the external inverse--Compton emission ($10^{19}$--$10^{20}$
         Hz). In the lower panel (d) we compare the 
         different types of the energy densities that control the particle 
         cooling. This shows that, in this particular simulation,
         the external inverse--Compton scattering 
         dominates the cooling process for large values of $\Gamma$ 
         whereas the synchrotron cooling is dominating for $\Gamma < 10$.
        }
\label{fig_mi1}        
\end{figure}

In Fig. \ref{fig_mi1} we compare the emission of different sources,
located at different distances from black hole that are traveling with
different velocities. In this test we assumed the following
physical parameters for the fastest source: 
$\Gamma_1=20$, $R'_1=2.4\times 10^{16}$ [cm], 
$B'_1=7$ [G], $\gamma_{i,1} = 6\times10^2$, 
$Q_{i,1} = 8 \times 10^{-5}$ [cm$^{-3}$ s$^{-1}$]. 
We calculate the parameters for the other 
sources with $\Gamma<\Gamma_1$ according to our prescriptions. 
Moreover, we assume that the sources are traveling within a constant 
radiation field, produced by a spherical shell 
($R_{\rm ext} = 3 \times 10^{17}$ [cm]). 
The emission of the shell is approximated 
as a black body with temperature $T_{\rm ext} = 2\times 10^4$ [K] 
and peak luminosity $\nu_{\rm max} L_{\rm ext} (\nu_{\rm max}) 
= 10^{45}$ [erg s$^{-1}$]. 
We assumed a redshift $z=0.5$, and the Hubble constant $H_0=72$ [km/(s Mpc)]. 
Since we will later apply our model to low energy peaked blazars 
the chosen parameters are somehow representative of these class of objects. 

The reason why we parametrize the fastest source and not the slowest 
one is purely ``technical". 
In order to calculate the particle distribution after 
the end of the injection we have to know the values of 
$U'_B$, $U'_{\rm syn}$ and $U'_{\rm ext}$. 
The values of
$U'_B$ and  $U'_{\rm ext}$ can be calculated straightforwardly from our 
parametrization, but the estimate of $U'_{\rm syn}$ is more complex,
since it depends on $N(\gamma)$, which in turn depends on $U'_{\rm syn}$.
We then proceed by calculating the spectra for sources with
different Lorentz factors in a sequence, decreasing the value
of $\Gamma$ by a very small factor. 
This allows us to use the value of $U'_{\rm syn}$ from the previous step 
in the current computations if the change of $U'_{\rm syn}$ from one step
to another is small. 
This method requires that in the first step the synchrotron radiation 
field energy density is negligible with respect to the other energy
densities. 
Since (for our set of parameters)
the fastest source cools mainly by the external radiation field, 
we always start the computations from the largest value of $\Gamma$.

Our set of input parameters implies $U'_{\rm ext} \gg U'_B \gg U'_{\rm syn}$ 
for the fastest source (Fig. \ref{fig_mi1} d). 
Therefore, in agreement with our analytical predictions, initially 
$F_{\rm EIC}(\nu)$ remains constant whereas $F_{\rm syn}(\nu)$ and 
$F_{\rm SSC}(\nu)$ 
are increasing fast for decreasing $\Gamma$ (Fig. \ref{fig_mi1} c). 
However, the decrease of the Lorentz factor quickly reduces the 
importance of the external radiation field in the comoving frame. 
Thus, the synchrotron emission becomes the dominant cooling process 
(Fig. \ref{fig_mi1} d). 
This leads to a
fast decrease of $F_{\rm EIC}(\nu)$ and also causes the decrease of 
$F_{\rm syn}(\nu)$ and $F_{\rm SSC}(\nu)$, as predicted 
(Fig. \ref{fig_mi1} c). 
Note that the parameters chosen for this particular simulation gives a
very weak contribution of the SSC process to the total 
emission that appears visible only in the sources with very 
small $\Gamma$ values (Fig.~\ref{fig_mi1} b). 

\begin{figure}[!t]
\resizebox{\hsize}{!}{\includegraphics{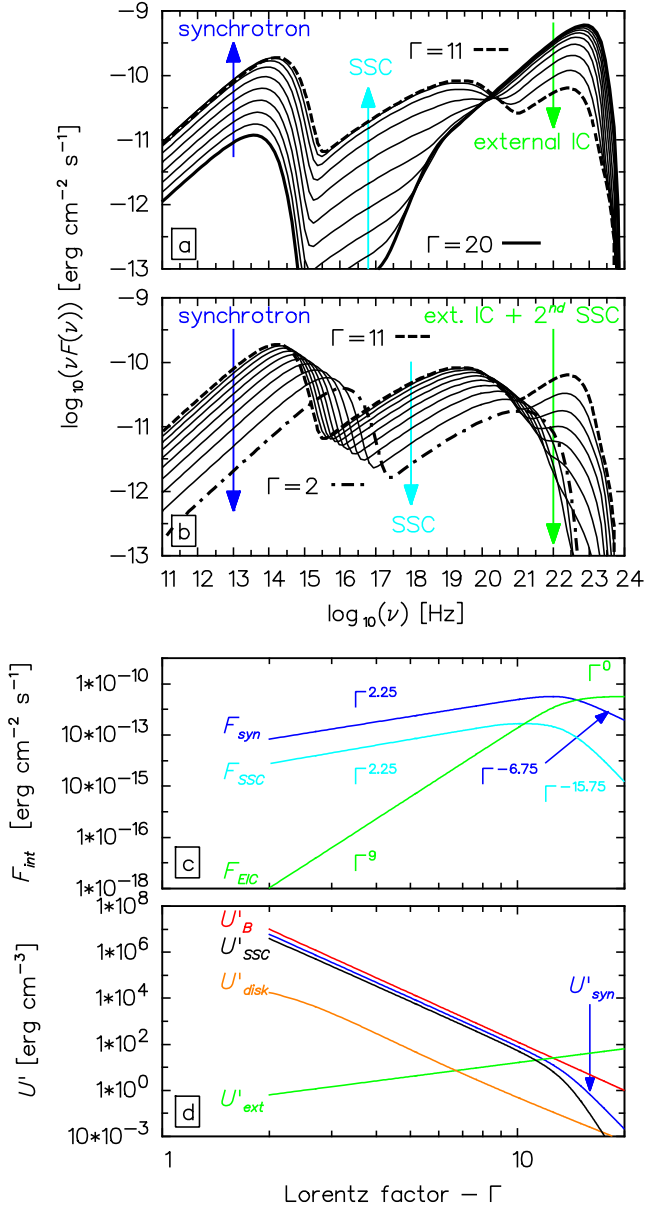}}
%\begin{figure}[!t]
%\includegraphics[height=19.5cm]{ps/art_mi2n_cp.ps} % fort the referee form
\caption{Top panels: comparison between spectra produced by different
         sources in the case of monoenergetic injection. In this
         particular test the SSC emission appears almost as
         important as the synchrotron radiation for sources with
         $\Gamma \sim 10$, for which the efficiency of
         the EIC process is significantly reduced. Note that 
         also the second order SSC process appears significant 
         in this particular case. The spectra of the different emission
         processes are presented in Fig. \ref{fig_mi2SED} for 
         three representative sources.
         Bottom panels: as in Fig. 1, it is shown how the fluxes
         of the different emission processes, and the corresponding 
         energy densities, scale as a function of $\Gamma$.
        }
\label{fig_mi2}        
\end{figure}

\begin{figure}[!t]
\resizebox{\hsize}{!}{\includegraphics{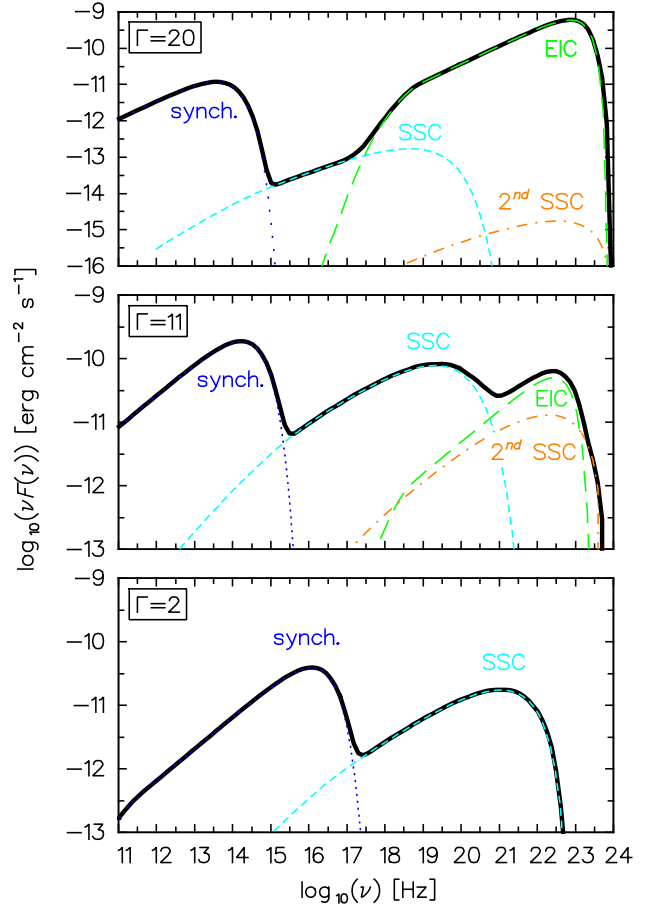}}
%\begin{figure}[!t]
%\includegraphics[height=19.5cm]{ps/art_mi2nSED_cp.ps} % fort the referee form
\caption{The components in the total spectrum produced by three different
         sources characterized by different values of $\Gamma$. The 
         spectra are chosen from the sequence presented in Fig. \ref{fig_mi2}.
         In the fastest source ($\Gamma=20$) the external inverse-Compton emission 
         (EIC, long dashed line) appears dominant. In the moderately fast source
         ($\Gamma=11$) the synchrotron (dotted line), SSC (short--dashed line) 
         and the EIC processes are producing similar amount of radiation.
         Moreover, also the second order SSC process (dash--dot line) is not
         negligible for this case. A relatively slow source ($\Gamma=2$) is 
         producing mostly synchrotron and SSC emission.
        }
\label{fig_mi2SED}        
\end{figure}

In the second test (Fig.~\ref{fig_mi2}) we modified some of the
physical parameters ($R'_1=2.4 \times 10^{16} \to 7.5 \times 10^{15}$ cm,
$B'=7 \to 5$ G, $Q_{i,1} = 8 \times 10^{-5} \to 1.5 \times 10^{-2}$ 
cm$^{-3}$ s$^{-1}$)  in order to amplify the SSC emission. 
However, 
even for this new set of the parameters the SSC and synchrotron 
cooling are negligible in the fastest source ($\Gamma_1=20$) where the
EIC emission dominates as in the previous test (Fig.~\ref{fig_mi2} a). 
As we already mentioned the EIC cooling 
decreases quickly with the decrease of $\Gamma$ at the advantage of 
other cooling processes. In this particular case, when $\Gamma\sim 10$,
the synchrotron and SSC cooling processes become dominant and 
almost equally important ($U'_B \sim U'_{\rm syn} \gg U'_{\rm ext}$, 
Fig.~\ref{fig_mi2} d). 
Moreover, also the second order SSC emission
appears significant in this particular test. 
However, the energy 
density of the SSC emission ($U'_{\rm SSC}$) that describes the
particle cooling due to 2$^{\rm nd}$ order SSC is for all the
sources a few times smaller than $U'_B$ (Fig.~\ref{fig_mi2} c). 
Note that if the SSC emission were stronger than the synchrotron 
radiation ($U'_{\rm syn} > U'_B$), the 
2$^{\rm nd}$ order SSC scattering would occur in the Klein--Nishina 
regime. 
This would require a special treatment of the cooling process 
and a numerical solution of the kinetic equation. 
Our parameter choice is such that $U'_{\rm syn} < U'_B$ in all cases.
As a consequence, the numerical calculations are always in perfect 
agreement with the analytic predictions. 
In Fig. \ref{fig_mi2SED} we show, separately, the spectra produced by
the different emission mechanisms, and how their importance changes
varying $\Gamma$.
The fastest source ($\Gamma=20$) is emitting the dominant part of 
the available energy through the EIC scattering. 
The moderately 
fast source ($\Gamma=11$) uses the synchrotron and the SSC 
processes to radiate the particle energy. 
The EIC in this source is significantly less important, producing a 
flux of the same order of the second order SSC emission. 
Finally,
the slowest source ($\Gamma=2$) is producing mostly
synchrotron and SSC radiation. 
The parameters of the sources span the ranges:  $R'=8 \times 10^{15} 
\to 8 \times 10^{13} $ cm, $B'=5 \to 1.6 \times 10^{4} $ G, and
$K=7.1 \times 10^{3} \to 2.3 \times 10^{7} $ cm. 

In all cases we calculate the optical depth for the
pair production process inside the source due to the interaction
between $\gamma$--rays and lower energy photons. 
This process appears negligible for all sources.

We estimated also the energy density of the radiation produced
by the accretion disk ($U'_{\rm disk}$) in the comoving
frame according to the prescription of Ghisellini \& Madau
(\cite{Ghisellini96}). 
For this estimate, we assume that
the total luminosity of the disk is ten times greater than 
the luminosity of the external radiation field, which is a 
good approximation if we assume that the external radiation 
field is produced by the broad line region. 
The distance between the putative black hole and the source ($D$) 
is calculated assuming that the jet is conical,
with an aperture angle of 0.1 rad ($\sim$5 degrees), 
giving  $D= R/0.1$.
The disk energy density starts to dominate over $U'_{\rm ext}$
at relatively small distances to the center, corresponding 
to small $\Gamma$--values ($ \Gamma \lesssim 5$).
Nevertheless, $U'_{\rm disk}$ remains always at least two
orders of magnitude weaker than the energy density of the
magnetic field and the internal radiation fields ($U'_{\rm syn}$ and 
$U'_{\rm SSC}$). Therefore, the disk radiation field is
negligible as a cooling agent of the particles inside the source.

The above tests show that according to our prescription the
fast moving sources radiate away the dissipated energy through
the inverse-Compton scattering off the external photons, whereas the 
slow moving sources may efficiently produce synchrotron and
also SSC emission. 
The relative level between the SSC and the synchrotron emission depends 
on the physical parameters of the source. 
What is important is that the same amount of the dissipated
energy is radiated away, but with different emission processes.
We can then observe completely different spectra originating
in the same jet, according to whether the dissipation takes place
in its inner (small $\Gamma$--values) or outer (large $\Gamma$) part.
For instance, the same cosmic source can display a state
with a moderate optical synchrotron 
emission and a strong $\gamma$--ray EIC radiation (at MeV--GeV energies), 
or a strong optical emission and a strong SSC radiation (at keV energies)
{\it without a corresponding higher level of the MeV--GeV flux.}
Dramatic variability of a particular object would not correspond
to dramatic variations on the jet efficiency (we have assumed that the blobs
always carry the same energy), but would rather reflect the place,
along the jet, where dissipation is taking place (if this location
is in turn linked with the bulk Lorentz factor).

Our model is rather ``economic" for the jet, in the sense
that the total energy radiated during different flares is approximately
the same, even if the variations in specific bands can be dramatic.
The fact that a synchrotron (in the IR--optical band) and an
X--ray flare may not be accompanied by a simultaneous flare
in the MeV--GeV band is the main prediction of our 
model, that can be easily verified by the coming new 
$\gamma$--ray observatories (GLAST, AGILE).
These kind of observations may prove if the jet is varying
its (likely kinetic, but possibly also magnetic) power or
if it is instead working at a constant average efficiency.

The monoenergetic injection used in our test provides a very 
simple description of the model and allows to describe the scaling
of different radiative processes in an analytic way.
However,
such simple injection is corresponding to a particle
distribution which is a power law, of slope $n=2$,
which cannot explain accurately most of the observations.
Therefore, in the next step we discuss the injection of a power
law particle distribution.

\subsection{Injection of a power law electron distribution}

In a homogeneous source, the spectrum of the observed emission 
depends directly on the shape of the particle distribution.
Therefore, to precisely reproduce the observed spectra 
of blazars, we need a particle energy spectrum that is 
more complex than a single power law provided by the 
monoenergetic injection. 
Quite a large flexibility in spectral fitting is obtained if we assume a 
power law energy distribution of the injected particles
\begin{equation}
Q(\gamma) = Q_i~ \gamma^{-n},~~ {\rm for}~~ 
\gamma_{\rm i} \le \gamma \le \gamma_{\rm max}.
\end{equation}
For $\gamma_{\rm i} > 1$ and $n>2$ such injection with the 
simultaneous radiative cooling provides a particle distribution 
that can be approximated by two power law functions
\begin{equation}
N(\gamma) = \left\{
             \begin{tabular}{ll}
              $K~ \gamma^{-2}$     & ${\rm for}~~ 
                 \gamma_{\rm min} \le \gamma \le \gamma_{\rm i}$ \\
              $K_2 \gamma^{-(n+1)}$ & ${\rm for}~~ 
                 \gamma_{\rm i}   <   \gamma \le \gamma_{\rm max}$
             \end{tabular}
            \right. ,
\end{equation}
where $K_2 = K \gamma_{i}^{n-1}$ and the particle density ($K$) is 
directly related to the injection rate ($Q_{\rm i}$) and the cooling
($C$) efficiency
\begin{equation}
K   = \frac{Q_{\rm i} }{C(n - 1) \gamma_{\rm i}^{n-1}},
\end{equation}
The minimum energy $\gamma_{\rm min}$ is calculated in the same way as for 
the monoenergetic injection (Eq. \ref{equ_mono_inj_elec_spec}) whereas  
$\gamma_{\rm max}$ is a free parameter, whose value is not important if $n> 2$. 
The injection rate can be calculated from the simple relation
\begin{equation}
Q_{\rm i}   = \frac{N_e}{T' V' (\gamma_{\rm max}^{1 - n} - \gamma_{\rm i}^{1 - n}) / (1 - n)},
\label{equ_pli_norm}
\end{equation}
where $N_e$ is the total number of injected particles.

We will assume that sources with different $\Gamma$--values may have
different slopes and different energy ranges of the injection,
but we still maintain our main assumption on the total energy of the injected 
electrons, which must be the same for all sources (Eq. \ref{equ_elec_energy}). 
This 
gives the following relation between the total number of particles, average
energy and $\Gamma$ in two different sources
\begin{equation}
N_{e,2}  = N_{e,1} \frac{\langle\gamma\rangle_1 \Gamma_1 }
{\langle\gamma\rangle_2 \Gamma_2}.
\end{equation}
\begin{figure*}[!t]
\resizebox{\hsize}{!}{\includegraphics{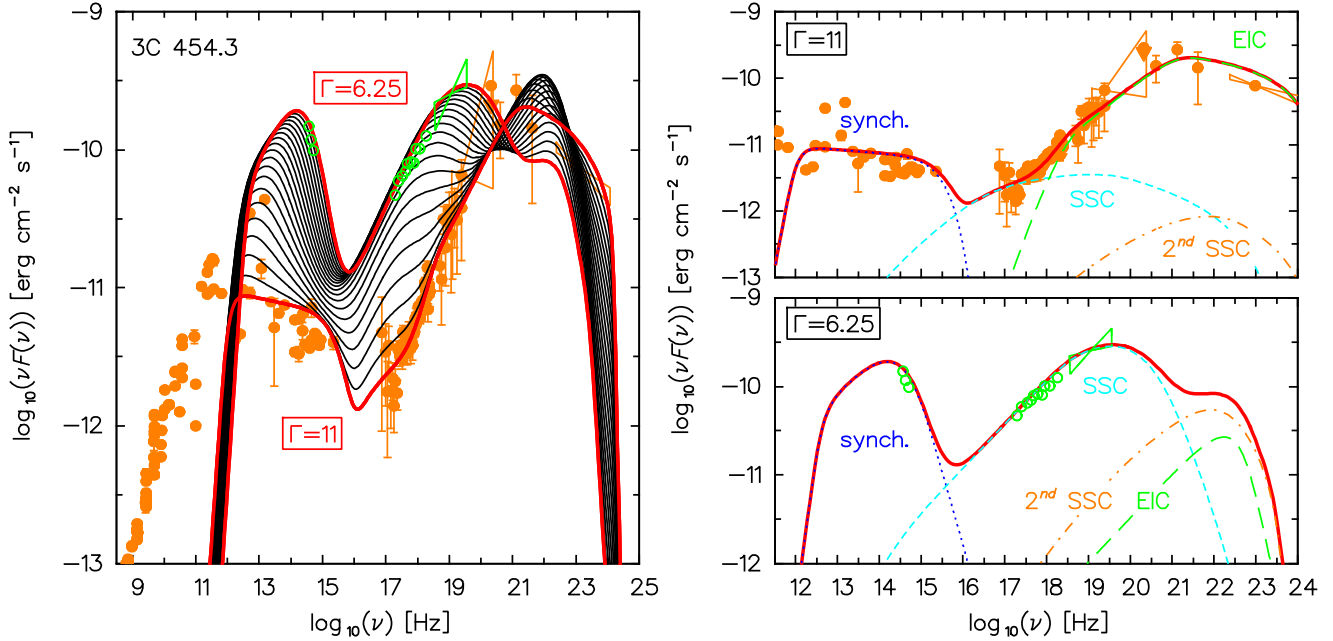}}
\caption{The multifrequency observations of 3C 454.3 for two different 
         emission states together with the results of our modelling. 
         The low level of the 
         optical and X--ray emission (state 1) is represented by data 
          (filled circles) 
         obtained at different epochs (references to the data 
         points in the text) whereas the high synchrotron and X--ray state
         (state 2) is represented by the observations 
         made by the Swift satellite and the REM telescope (Giommi et al. 
         \cite{Giommi06}).
         The left panel shows a sequence of spectra with
         different Lorentz factors: the model with the largest $\Gamma$ 
         reproduces state 1 and the model with the smallest $\Gamma$ 
         can explain state 2. In the right panels we show
         how the different emission processes contribute to the total spectrum.
        }
\label{fig_pliSED}        
\end{figure*}

Using the above description we can calculate the emission from different sources
assuming different injection profiles but using the same amount of the
injected energy. 
Even if the power law injection is more realistic than the 
monoenergetic injection, the results obtained
are qualitatively the same for both cases.

We will directly apply our simulations, done for a power law injection 
case, to the recent observations of the blazar 3C 454.3, which we will
use as an illustrative example.

\section{Application to 3C 454.3}

Recent observations of the blazar 3C 454.3 showed a huge flare 
in the optical and
X--ray bands (Fuhrmann et al. \cite{Fuhrmann06}, Pian et al. \cite{Pian06},
Villata et al. \cite{Villata06}, Giommi et al. \cite{Giommi06}). 
We will investigate if such activity
is in agreement with the prediction of our model, in which 
sources with relatively small velocities ($\Gamma \lesssim 10$) can produce 
mostly synchrotron (IR $\to$ UV) and SSC (UV $\to$ hard X--ray) emission.
Therefore in the next step we will try to reproduce this unusual activity.

In Fig. \ref{fig_pliSED} we collected the multifrequency observations of
3C~454.3. 
Most of the data are historical observations made at different 
epochs, illustrating the low level of the optical and X--ray emission.
The historical data from the radio frequencies up to the optical range
are taken mostly from the NED database. 
However, in this range we are
plotting also a few points from other sources (see Pian et al. 
\cite{Pian06} and references therein). 
In the X--ray range we show
the low/hard emission level observed by $Beppo$SAX in 2000 
(Tavecchio et al. \cite{Tavecchio02}). 
Finally, in hard X--rays and $\gamma$--rays we show the
COMPTEL and EGRET measurements made in the years 1991--1994 (Zhang et al.
\cite{Zhang05}, Hartman et al. \cite{Hartman99}). 
The high level of
the optical and X--ray emission is shown by the data taken
simultaneously by three experiments (UVOT, XRT, BAT) onboard the Swift
satellite and by the quasi--simultaneous ground--based data taken by the
REM telescope (Giommi et al. \cite{Giommi06}).
We will focus onto two states of the source:
state 1 corresponds to the high level of $\gamma$--ray emission,
low/hard X--ray flux and low IR--optical flux, 
while state 2 corresponds to the recent optical and X--ray flare.
Note that position of the synchrotron and the IC peak as well, 
is not well determined for the second state.

We first try to reproduce state 1.
Assuming that the $\gamma$--rays are produced by the EIC scattering inside
a relatively fast moving source, we obtained the best fit using 
the following parameters: 
$\Gamma_1=\delta_1=11$, 
$R'_1=1.1 \times 10^{16}$ cm, $B'_1=4.8$ G, 
$\gamma_{\rm i, 1} = 90$, $\gamma_{\rm max} = 5 \times 10^3$, 
$Q_{\rm i,1} = 25.4$ cm$^{-3}$ s$^{-1}$, $n_1=2.1$.
For the external radiation field we used 
$\nu_{\rm max} L_{\rm ext}(\nu_{\rm max}) = 10^{45}$ erg s$^{-1}$, 
$T_{\rm ext}=2 \times 10^4$ K, 
$R_{\rm ext} = 2.8 \times 10^{17}$ cm. 
The spectral index of the injected particles is relatively soft ($n_1=2.1$), to
explain the soft spectrum observed in the IR--UV range and also the
relatively soft spectrum in the MeV--GeV range observed by EGRET.

In order to reproduce state 2 we calculate a
sequence of spectra decreasing the value of $\Gamma$ 
($\Gamma=11 \to \Gamma=6.25$) by steps of $\Delta \Gamma=0.25$.
Moreover, in each step we increase the spectral 
index of the injected particle distribution by a factor $\Delta n=1/3$. 
Also the minimum energy of the injected particles $\gamma_{\rm i}$ increases
by a factor $\Delta \gamma_{\rm i} = 0.6 \gamma_{\rm i,1}$ 
whereas $\gamma_{\rm max}$ is assumed to be constant.
According to our main assumption the changes of the injection spectrum 
are compensated by the change of the normalization (Eq. \ref{equ_pli_norm}), 
to provide the same amount of the injected energy for all states.
Note that in this case we maintain the viewing angle fixed
($\theta=5.2^\circ$) and therefore, in general, $\delta\ne \Gamma$. 
The best fit for state 2 is reached for 
$\Gamma_2=6.25$, $\delta_2=9.42$, 
$R'_2=3.7 \times 10^{15}$ cm, $B'_2=35.7$ G, 
$\gamma_{\rm i, 2} = 6 \times 10^2$, 
$Q_{\rm i,2} = 3.4 \times 10^{13}$ cm$^{-3}$ s$^{-1}$, $n_2=5.26$.
Note that while the set of parameters used for state 1 was chosen
to best represent the data, the parameters for state 2 are given
by our assumed parametrization, and are not free.
Note also that we here appropriately calculate the  synchrotron self--absorbed 
spectra, that we neglected previously.

Self--absorption truncates the synchrotron spectrum below the IR range. 
This effect appears frequently in the modelling of the high
energy emission of blazars and is related to the size of the source
that must be relatively small ($10^{15\to16}$ [cm]) in order to 
explain the observed variability time scales.
Such a small source must
be also relatively dense (both in particles and magnetic field)
to explain the observed emission level and therefore it is 
optically thick at radio frequencies. 
The radio emission should be produced by more extended radio
structures, well visible in the VLBI maps. 
Moreover, this particular blazar does not show variability of the 
radio flux clearly correlated with the optical, X--ray and 
$\gamma$--ray activity (e.g. Villata et al. \cite{Villata06}). 
This suggests that the quiescent radio and the variable 
optical--to--$\gamma$--ray emission are produced in different regions 
of the source. 
At any rate, the analysis of the radio emission produced probably 
by an inhomogeneous jet is out of the scope of this work.

Note that our model reproducing state 2 is very similar to the
spectrum resulting from the modelling (of the same state) 
made by Pian et al.~(\cite{Pian06}).
In that paper, the emitting region was  assumed to be
outside the broad line region, in order to neglect the 
external radiation and the corresponding external Compton 
emission.
In order to compensate the relatively low values of the magnetic field and
the particle density, appropriate at such large distances from the center,
it was necessary to adopt a large $\Gamma$, giving
a Doppler factor $\delta=15$.
This difference in the $\Gamma$--factors gives a potential tool to 
discriminate between the model presented here (which has $\delta=9.4$)
and the model discussed in Pian et al. (\cite{Pian06}).
In fact, in our model, the apparent velocities of VLBI knots (after an 
optical and X--ray flare) should be smaller than in the Pian et al. model.
Unfortunately, the predicted differences are not large, and probably
not measurable with enough accuracy for this particular blazar.

The presented simulation shows that our approach can well explain
two different levels of the emission observed in 3C~454.3. 
We stress that what we proposed is the most ``economic'' 
way to reproduce dramatically different levels of emission, even 
if the jet is using always the same amount of energy.
 
The strong prediction of this approach is that the level of the 
$\gamma$--ray MeV--GeV emission should be lower (or not particularly 
bright) when the level of the IR--to--X--ray emission is high. 
For the moment there is no evidence for such behaviour. 
We must wait for simultaneous optical, X--ray and especially
$\gamma$--ray observations of blazars to verify if our 
idea is correct.

\section{Summary and conclusions}

We have proposed a model to explain different levels of 
the emission observed from the IR range up to the $\gamma$--ray band 
in a powerful blazars. 
The main assumptions of our model is that the emission zone of the jet 
is dissipating always the same amount of energy, but that the emission 
zone can be located in different parts of the jet, farther from the 
central engine if the bulk Lorentz factor is larger.
We have discussed a particular scenario where this naturally occurs,
namely the internal shock scenario.
While this scenario is not mandatory to develop our main idea,
it is the only existing scenario where we can work out how all 
the relevant quantities (emission volume, magnetic field, particle 
density and so on) scale with the bulk Lorentz factor $\Gamma$.
The obtained results are certainly promising, since even with
our simplifying assumption (the emission sites always dissipate
the same amount of total energy) we were able to explain
dramatically different states observed at different frequency bands.
In particular, we were able to explain two very different states of the 
blazar 3C 454.3.
This is achieved by noting that when the bulk Lorentz factor $\Gamma$ 
is small, the shell--shell collision in the internal shock scenario occur
closer to the apex of the jet, therefore in a more compact region,
with a larger magnetic field. 
Since the external Compton radiation is dimmed in this case 
(being less boosted in the comoving frame), the electron energy
is radiated more through the synchrotron and the self--Compton
processes, therefore not at MeV--GeV energies, but in the IR--optical
(through synchrotron) and X--rays (through self--Compton).
Lowering the $\Gamma$--factor there is therefore a ``transfer of power"
from higher to lower frequencies.

If this idea is correct, there should exist sources flaring in the
optical and X--ray bands, but not in $\gamma$--rays, where the flux
can even decrease.
This prediction can be easily tested by the forthcoming
$\gamma$--ray satellites, AGILE and GLAST, together with
simultaneous optical and X--ray observations: 
had MeV-GeV observations been available
during the 2005 flare of 3C 454.3,
such a test could have already been done.

The possibility that the jet can dissipate part of its bulk
kinetic energy at small distances from its apex, and at small
distances from the accretion disk, may at first sight contradict
what claimed in Ghisellini \& Madau (1996), where strong
dissipation close to disk was excluded.
But in that case ``dissipation" meant strong production of
$\gamma$--rays, dominating the radiative output.
Here, instead, when the dissipation occurs ``early"
(i.e. close to the disk) the external Compton flux is unimportant,
and the emission at high energies, made by the second order
self--Compton emission, is never dominant.
Therefore, even if a fraction of $\gamma$--rays get absorbed
in $\gamma$--$\gamma$ collision producing pairs (likely with the
X--ray emission from the disk corona),
the spectrum is not strongly reprocessed, and the emission from 
the newly created pairs is not important enough to ``fill the valley"
between the two broad peaks of the blazar SED.

We do not know the relative fraction of time a jet
spends producing blobs with large and small bulk
Lorentz factors.
Consider also that our current knowledge of blazars
may be biased by the blazars observed (and detected)
by EGRET.
But, as mentioned in the introduction, EGRET detected
only one fourth of the  radio brightest blazars.
This may suggest that a small value of the bulk Lorentz
factor is the rule, not the exception, since in this
regime the synchrotron emission is favoured with respect to
the external Compton emission.
The factor $\sim$20 better sensitivity of GLAST
with respect to EGRET will surely be crucial to solve this
issue, and to shed light on these particular aspects of
powerful relativistic jet.

\begin{acknowledgements}
This research has made use of the NASA/IPAC Extragalactic Database 
(NED) which is operated by the Jet Propulsion Laboratory, California 
Institute of Technology, under contract with the National Aeronautics 
and Space Administration. We acknowledge the EC funding under contract 
HPRCN-CT-2002-00321 (ENIGMA network).
\end{acknowledgements}

\end{document}